\documentstyle[12pt]{article}
\title{Gravitational crystal inside the black hole}
\author{Hrvoje Nikoli\'c \\
Theoretical Physics Division, Rudjer Bo\v{s}kovi\'{c} Institute, \\
P.O.B. 180, HR-10002 Zagreb, Croatia \\
{\normalsize e-mail: hnikolic@irb.hr} \\
\makebox[1in]{} \\
}
\date{\today}
\begin{document}
\maketitle
\begin{abstract}
Crystals, as quantum objects typically much larger than their lattice spacing, 
are a counterexample to a frequent prejudice that quantum effects should not be 
pronounced at macroscopic distances.
We propose that
the Einstein theory of gravity only describes a fluid phase 
and that a phase transition of crystallization can occur under extreme conditions such as those inside
the black hole. Such a crystal phase with lattice spacing of the order of the Planck length
offers a natural mechanism for pronounced quantum-gravity effects at distances much larger than the Planck length.
A resolution of the black-hole information paradox is proposed, according to which
all information is stored in a crystal-phase remnant with size and mass much above the Planck scale.
\end{abstract}
\vspace*{0.5cm}

\section{Introduction and main ideas}

The black-hole information paradox (see e.g. \cite{gid,har,pres,pag,gid2,str,math,hoss,harlow,fabbri} for reviews)
is one of the greatest challenges for a successful marriage between theory of gravity and quantum theory.
Many proposals for the resolution of the paradox, such as massive black-hole remnant \cite{gid},
fuzzball \cite{mathur}, energetic curtain \cite{braunstein}, firewall \cite{AMPS}, or Planck star \cite{rovelli},
require a large deviation from classical gravity at macroscopic distances.  
On the other hand, classical gravity is generally expected to be a very good approximation at macroscopic distances,
which raises justified skepticism about all such proposals of large deviations from classical gravity 
at macroscopic distances.

Nevertheless, it is not true that quantum effects cannot be significant at macroscopic distances.
At low temperatures, some materials show macroscopic features such as superfluidity or superconductivity
\cite{annett}, which cannot be explained by classical physics. Furthermore, there are even examples which
do not require low temperature.
For instance, white dwarf stars avoid gravitational collapse owing to the electron degeneracy pressure 
\cite{perkins}, which is a consequence of the quantum Fermi-Dirac statistics.
Similarly, neutron stars avoid gravitational collapse owing to the quantum degeneracy pressure of neutrons \cite{perkins}.

Finally, even if these examples look somewhat exotic, there is a totally non-exotic example; 
probably the best understood and most ubiquitous macroscopic quantum object is a {\em crystal} \cite{marder}.
Crystals can be found everywhere in our everyday macroscopic lives. 
And yet, they are quantum objects which can be thought of
as macroscopic ``molecules''. Indeed, the typical lattice spacing between the neighboring atoms in a crystal 
is a few Bohr radii, where the Bohr radius (in Gaussian CGS units) 
\begin{equation}\label{bohr}
a_0= \frac{\hbar^2}{me^2}
\end{equation}
is a quantum distance, determined by the Planck constant $\hbar$, electron mass $m$, and electron charge $e$.
The lattice spacing in the crystal determines many of its macroscopic properties. In particular, the fact that 
spacing is of the order of $a_0$ implies that entropy density $s$ in a crystal is of the order  
\begin{equation}\label{s}
 s\sim \frac{1}{a_0^3} .
\end{equation}

The analog of (\ref{bohr}) in quantum gravity is the Planck length $l_{\rm Pl}$, given by
\begin{equation}\label{planck}
l_{\rm Pl}= \frac{1}{m_{\rm Pl}}
\end{equation}
in the units $\hbar=c=1$, where $m_{\rm Pl}$ is the Planck mass. 
It is often assumed in the literature that effects of quantum gravity 
should not be seen at distances much larger than $l_{\rm Pl}$. Yet, the analogy between 
(\ref{bohr}) and (\ref{planck}) suggests that such an assumption may not be justified.
Just as the Bohr radius has a macroscopic manifestation in matter crystals, the Planck length
could have similar macroscopic manifestation in a {\em gravitational crystal}. In particular,
the analogy with (\ref{s}) suggests that 
the entropy density of such a gravitational crystal could be of the order of 
the Planckian entropy density
\begin{equation}\label{planck-dens}
s\sim s_{\rm Pl}\equiv \frac{1}{l_{\rm Pl}^3} .
\end{equation}

Can a gravitational crystal be described by the Einstein-Hilbert action of gravity?
The classical and semi-classical gravitational phenomena have many similarities 
with phenomena in condensed-matter physics \cite{volovik,NVV,BLV},
but most of these condensed-matter phenomena are properties of fluid phases,
while similarities between gravity and the properties of crystals are not so pronounced.
Guided by general principles of condensed matter physics \cite{marder}, we suggest
that this is because the classical Einstein theory of gravity is really an effective macroscopic 
description of a fluid. In other words, we propose that the classical Einstein equation  
only describes the {\em fluid phase} of some unknown fundamental degrees of freedom. 
If so, then {\em the crystal phase cannot be described by quantization of the Einstein-Hilbert action},
implying that the Einstein-Hilbert action is not fundamental.
Roughly, this is analogous to the fact that ice cannot be described by quantization of a
non-fundamental macroscopic fluid equation, such as the Euler equation or the Navier-Stokes equation. 
The transition from the fluid phase (described by general relativity) to the crystal phase 
(described by as yet unknown theory)
occurs via a phase transition (for a related idea see also \cite{chapline}). 
Unfortunately, in the absence of detailed knowledge about the fundamental theory, 
the details of such a phase transition cannot be described explicitly from first principles.

\section{A simple model}

Even though we cannot derive the properties of such a gravitational crystal from first principles,
some essential qualitative properties can be described by a simple model.
For that purpose, we assume that crystallization happens under extreme conditions inside the black hole, 
in the core formed around the center at $r=0$. 
Denoting by $r_{\rm core}$ the radius of the crystallized core, general relativity
is valid only in the fluid phase at $r>r_{\rm core}$. The physics in the core 
for $r\leq r_{\rm core}$ is not described by the Einstein-Hilbert action, 
and {\it a priori} the entropy of the
core does not need to scale with area. Therefore, to model the core, we take the simplest possible assumption,
i.e. we assume that the effective metric in the core is the flat Minkowski metric and that 
its entropy scales with volume in accordance with (\ref{planck-dens}).
(The flat metric in the crystal phase seems natural from a condensed-matter point of view because,
in condensed-matter physics, the fundamental degrees of freedom do not describe metric.
The existence of a curved metric as an effective large-distance description is characteristic 
for the fluid phase \cite{NVV,BLV}, but not for the crystal phase.)
Hence the volume of the core is $V_{\rm core}=4\pi r_{\rm core}^3/3$ and its entropy is
\begin{equation}\label{Score}
 S_{\rm core}=\alpha V_{\rm core} = \frac{4\pi\alpha}{3}r_{\rm core}^3 ,
\end{equation}
where $\alpha\sim 1$ and we work in Planckian units $l_{\rm Pl}=m_{\rm Pl}=1$. 
(For a related model of a core inside the black hole see also \cite{nicolini}.)
On the other hand, the fluid phase outside the core is described by general relativity,
so the fluid-phase entropy is the standard Bekenstein-Hawking entropy
\begin{equation}\label{SBH}
 S_{\rm BH}=\frac{A}{4}=\pi R^2 =4\pi M^2,
\end{equation}
where $A=4\pi R^2$ is the black-hole area, $R=2M$ is the black-hole radius, 
and $M$ is the black-hole mass.
The total black-hole entropy $S_{\rm bh}$ is the sum of the two contributions
\begin{equation}\label{Sbh1}
 S_{\rm bh}=S_{\rm BH}+S_{\rm core}.
\end{equation}

Now let us assume that initially there is no core, so that the initial black-hole entropy 
is the initial Bekenstein-Hawking entropy
\begin{equation}\label{S0}
 S_0=\frac{A_0}{4}=4\pi M_0^2 ,
\end{equation}
where $M_0$ is the initial black-hole mass. Owing to the Hawking radiation \cite{hawk1}, the mass $M$ 
decreases and the black hole absorbs entropy of the ingoing Hawking quanta. This entropy cannot
be absorbed by the fluid phase because the fluid-phase entropy $A/4$ decreases due to the decrease of $M$.
Consequently, a crystal phase must be created in the center of the black hole 
in order to absorb the incoming entropy. The incoming entropy is the entropy of entanglement
with external Hawking radiation, so the incoming entropy is equal to the radiation entropy
$S_{\rm radiation}$. Hence the total black-hole entropy must be equal to
\begin{equation}\label{Sbh2}
 S_{\rm bh}=S_0+S_{\rm radiation}.
\end{equation} 

The radiation entropy has been calculated by Page \cite{page1,page2,page3}. His result can be written as
\begin{equation}\label{Srad}
 S_{\rm radiation}=\eta\left( \frac{A_0}{4}-\frac{A}{4} \right) = \eta(4\pi M_0^2-4\pi M^2) ,
\end{equation}
where $\eta\approx 1.5$. 

Equating (\ref{Sbh1}) with (\ref{Sbh2}) and using (\ref{Score}), (\ref{SBH}), (\ref{S0}) and 
(\ref{Srad}), we obtain 
\begin{equation}\label{evol}
\frac{\alpha}{3} r_{\rm core}^3 = (1+\eta) (M_0^2-M^2),
\end{equation}
which describes how $r_{\rm core}$ increases as the black-hole mass $M$ decreases due to Hawking radiation.
As the black-hole radius $R=2M$ shrinks, the crystal radius $r_{\rm core}$ grows.

This black-hole shrinking and crystal growth is a continuous process which lasts as long as 
Hawking radiation is created at the horizon at $R=2M$. This happens as long as general relativity 
is valid at $r\geq R$. However, at some point $r_{\rm core}$ becomes equal to $R$, at which point
general relativity ceases to be valid at the horizon. At this point there is no reason to expect
any further creation of Hawking radiation, so the process of crystal growth stops at that point.
At this critical point we have $r_{\rm core}=R=2M$, so (\ref{evol}) gives
\begin{equation}\label{cubic}
 \frac{8\alpha}{3} M^3 = (1+\eta) (M_0^2-M^2) . 
\end{equation}
which is a cubic equation for $M$. 

The exact solution of this cubic equation can be expressed in an analytic form \cite{bronshtein},
but such an expression is rather cumbersome and not very illuminating. It is much more illuminating
to find an approximate analytic solution of (\ref{cubic}). Since $\alpha$ and $\eta$ are of the order of unity,
while $M_0\gg 1$, it is not difficult to see that the solution of (\ref{cubic}) satisfies
$M\ll M_0$. Hence (\ref{cubic}) can be approximated by the equation
\begin{equation}\label{cubic2}
 \frac{8\alpha}{3} M^3 = (1+\eta) M_0^2 , 
\end{equation}
the analytic solution of which has a very simple form
\begin{equation}\label{M}
M=\gamma M_0^{2/3} ,
\end{equation}
with
\begin{equation}
\gamma\equiv \left( \frac{3(1+\eta)}{8\alpha} \right)^{1/3} .
\end{equation}

To better understand the physical content of (\ref{M}), it is useful to restore the physical units in which 
$m_{\rm Pl}\neq 1$. In such units, (\ref{M}) can be written as
\begin{equation}\label{M2}
M=\gamma\, m_{\rm Pl} \left( \frac{M_0}{m_{\rm Pl}} \right)^{2/3} 
= \gamma M_0 \left( \frac{m_{\rm Pl}}{M_0} \right)^{1/3} .
\end{equation}
Since $M_0\gg m_{\rm Pl}$, this implies a hierarchy
\begin{equation}
 m_{\rm Pl}\ll M \ll M_0 .
\end{equation}
The mass $M$ given by (\ref{M2}) is the mass of the black-hole remnant, the mass and size of which 
are much above the Planck scale.

As a possible resolution of the black-hole information paradox,
a massive remnant with mass of the order of (\ref{M2}) has also been suggested by Giddings
in \cite{gid}. The problem with his suggestion, in his own words, was the acausal behavior of the 
core behind the horizon. Namely, if one assumes that general relativity is valid in the core, 
then causal evolution of the core is not compatible with an increasing core radius $r_{\rm core}$.
But in our approach this is not a problem, because the core is in the crystal phase,
while general relativity is valid only in the fluid phase outside the core.

\section{Discussion}

In the literature one can find many other proposals for the black-hole remnant (see \cite{chen} for a review),
most of which have mass on the Planck scale or some other fixed scale which (unlike (\ref{M2})) does not
depend on $M_0$. To resolve the black-hole information paradox for an arbitrarily large $M_0$, 
such a remnant should be able to absorb an arbitrarily large amount of information.
But object with a bounded mass and unbounded phase space is expected 
to have an unbounded probability of production in various quantum processes \cite{gid,har,pres,pag,gid2,str}, 
contradicting the fact that such productions are not observed.
There are also various counterarguments \cite{chen} according to which the overproduction of such objects
should not necessarily be expected. But whatever one thinks of such counterarguments, 
we emphasize that in our scenario the mass of the remnant (\ref{M2}) increases with  $M_0$,
so our remnant does not lead to such a problem at all.

In our scenario the entropy of the remnant is proportional to the volume rather than the area.
At first sight, this may seem to be incompatible with AdS/CFT duality \cite{maldacena,natsuume}. 
Nevertheless, there are two possibilities for making it compatible with AdS/CFT.
The first possibility is that AdS/CFT duality is a property of an approximative theory
(for example, it is possible that string theory is only an approximative theory of quantum gravity),
while the gravitational crystal is described by some more fundamental, as yet unknown theory.
The second, more interesting possibility, is that AdS/CFT itself allows entropy which scales with volume,
as proposed by the non-holographic version of AdS/CFT \cite{nik-adscft}.      

\section{Conclusion}

Inspired by insights from condensed matter physics, 
in this paper we have considered a possibility that general relativity is only an effective large-distance
description of a fluid phase of some unknown fundamental degrees of freedom.
From that point of view, it seems natural to assume that the same fundamental degrees of freedom
can also exist in the crystal phase, and that entropy of the crystal phase can scale with
volume rather than area. Under these assumptions, we have proposed a natural resolution 
of the black-hole information paradox, according to which all the information needed for unitarity of Hawking radiation
is absorbed by the crystal core inside the black hole. Even if there is
no such a core in the initial black-hole state, the creation of a core (via a phase transition of crystallization)
is forced by incoming information of ingoing Hawking quanta. This provides a simple mechanism for crystal growth,
which eventually leads to the result that the final state of black-hole evaporation is a remnant
in the crystal phase, with mass and size much above the Planck scale.
Even though such a large remnant requires effects of quantum gravity at macroscopic distances, 
the idea that these effects are a consequence of crystallization makes such effects rather natural;
the violation of classical general relativity in a macroscopic gravitational crystal should not be 
more ``unexpected'' than the violation of the classical Navier-Stokes fluid equation in a macroscopic piece
of ice.
Hence the gravitational crystal inside the black hole, as a candidate for a
massive remnant which stores information and resolves the black-hole information 
paradox, seems to be an attractive idea worthwhile of further investigation.

\section*{Acknowledgements}

This work was supported by the Ministry of Science of the Republic of Croatia.


\begin{thebibliography}{99}

 

\bibitem{gid}
S.B. Giddings, Phys. Rev. D {\bf 46}, 1347 (1992); hep-th/9203059. 
\bibitem{har} 
J.A.~Harvey, A.~Strominger, hep-th/9209055.
\bibitem{pres}
J. Preskill, hep-th/9209058.
\bibitem{pag} 
D.N.~Page, hep-th/9305040.
\bibitem{gid2} 
S.B.~Giddings, hep-th/9412138.
\bibitem{str}
A.~Strominger, hep-th/9501071.
\bibitem{math}
S.D. Mathur, Lect. Notes Phys. {\bf 769}, 3 (2009); arXiv:0803.2030.
\bibitem{hoss}
S. Hossenfelder, L. Smolin, Phys. Rev. D {\bf 81}, 064009 (2010); arXiv:0901.3156.
\bibitem{harlow}
D. Harlow, arXiv:1409.1231.
\bibitem{fabbri}
A. Fabbri, J. Navarro-Salas, {\it Modeling Black Hole Evaporation}
(Imperial College Press, London, 2005).

\bibitem{mathur}
S.D. Mathur, Fortsch. Phys. {\bf 53}, 793 (2005); hep-th/0502050.
\bibitem{braunstein}
S.L. Braunstein, S. Pirandola, K. Zyczkowski, 
Phys. Rev. Lett. {\bf 110}, 101301 (2013); arXiv:0907.1190.
\bibitem{AMPS}
A. Almheiri, D. Marolf, J. Polchinski, J. Sully, JHEP {\bf 1302}, 062 (2013); arXiv:1207.3123.
\bibitem{rovelli}
C. Rovelli, F. Vidotto, arXiv:1401.6562.

\bibitem{annett}
J.M. Annett, {\it Superconductivity, Superfluids, and Condensates} 
(Oxford University Press, Oxford, 2004).

\bibitem{perkins}
D.H. Perkins, {\it Particle Astrophysics} (Oxford University Press, Oxford, 2009).

\bibitem{marder}
M.P. Marder, {\it Condensed Matter Physics} (John Wiley \& Sons, New Jersey, 2010). 

\bibitem{volovik}
G.E. Volovik, Phys. Rep. {\bf 351}, 195 (2001); gr-qc/0005091.
\bibitem{NVV}
M. Novello, M. Visser, G. Volovik (eds), {\it Artificial Black Holes}
(World Scientific, New Jersey, 2002).
\bibitem{BLV}
C. Barcelo, S. Liberati, M. Visser, Living Rev. Rel. {\bf 8}, 12 (2005); gr-qc/0505065.



\bibitem{chapline}
G. Chapline, E. Hohlfeld, R.B. Laughlin, D.I. Santiago, 
Int. J. Mod. Phys. A {\bf 18}, 3587 (2003); gr-qc/0012094.

\bibitem{nicolini}
P. Nicolini, D. Singleton, Phys. Lett. B {\bf 738}, 213 (2014); arXiv:1409.5069.

\bibitem{hawk1}
S.W.~Hawking, Commun.~Math.~Phys.~{\bf 43}, 199 (1975).

\bibitem{page1}
D.N.~Page, Phys. Rev. Lett. {\bf 50}, 1013 (1983). 
\bibitem{page2}
D.N.~Page, New J. Phys. {\bf 7}, 203 (2005); hep-th/0409024. 
\bibitem{page3}
D.N.~Page, JCAP {\bf 1309}, 028 (2013); arXiv:1301.4995.

\bibitem{bronshtein}
I.N. Bronshtein et al, {\it Handbook of Mathematics} 
(Springer-Verlag, Berlin, 2007).

\bibitem{chen}
P. Chen, Y.C. Ong, D.-h. Yeom, arXiv:1412.8366.

\bibitem{maldacena}
J.M. Maldacena, Adv. Theor. Math. Phys. {\bf 2}, 231 (1998); hep-th/9711200
\bibitem{natsuume} 
M. Natsuume, {\it AdS/CFT Duality User Guide} (Springer, Tokyo, 2015); arXiv:1409.3575.

\bibitem{nik-adscft}
H. Nikoli\'c, arXiv:1507.00591.

\end{thebibliography}
\end{document}